\documentclass[aps,pra,twocolumn,noshowpacs,superscriptaddress]{revtex4-1}
\usepackage{graphicx,epsfig, subfigure}
\usepackage{amsbsy}
\usepackage{amssymb}
\usepackage{amsmath}
\usepackage{physics}
\usepackage{hyperref}
\usepackage[bottom]{footmisc}
\usepackage{xcolor}
\usepackage{color}
\usepackage{times}
\usepackage{textcomp}
\usepackage{verbatim} 

\date{\today}

\begin{document}

\newcommand{\eqnref}[1]{Eq.~\ref{#1}}
\newcommand{\figref}[2][]{Fig.~\ref{#2}#1}
\newcommand{\RN}[1]{%
  \textup{\uppercase\expandafter{\romannumeral#1}}%
}

\title{Stable emission and fast optical modulation of quantum emitters in boron nitride nanotubes}

\author{Jonghoon Ahn}
\thanks{Equal contribution}
\affiliation{School of Electrical and Computer Engineering, Purdue University, West Lafayette, Indiana 47907, USA}

\author{Zhujing Xu}
\thanks{Equal contribution}
	\affiliation{Department of Physics and Astronomy, Purdue University, West Lafayette, Indiana 47907, USA}

\author{Jaehoon Bang}
\affiliation{School of Electrical and Computer Engineering, Purdue University, West Lafayette, Indiana 47907, USA}

\author{Andres E. Llacsahuanga Allcca}
	\affiliation{Department of Physics and Astronomy, Purdue University, West Lafayette, Indiana 47907, USA}

\author{Yong P. Chen}
\affiliation{School of Electrical and Computer Engineering, Purdue University, West Lafayette, Indiana 47907, USA}
	\affiliation{Department of Physics and Astronomy, Purdue University, West Lafayette, Indiana 47907, USA}
	\affiliation{Birck Nanotechnology Center, Purdue University, West Lafayette, Indiana 47907, USA}
	\affiliation{Purdue Quantum Center, Purdue University, West Lafayette, Indiana 47907, USA}

\author{Tongcang Li}
	\email{tcli@purdue.edu}
	\affiliation{School of Electrical and Computer Engineering, Purdue University, West Lafayette, Indiana 47907, USA}
	\affiliation{Department of Physics and Astronomy, Purdue University, West Lafayette, Indiana 47907, USA}
	\affiliation{Birck Nanotechnology Center, Purdue University, West Lafayette, Indiana 47907, USA}
	\affiliation{Purdue Quantum Center, Purdue University, West Lafayette, Indiana 47907, USA}	
	\date{\today}

\begin{abstract}

Atom-like defects in two-dimensional (2D) hexagonal boron nitride (hBN) have recently emerged as a promising platform for quantum information science. Here we investigate single-photon emissions from atomic defects in boron nitride nanotubes (BNNTs).  We demonstrate the first optical modulation of the quantum emission from BNNTs with a near-infrared laser. This one-dimensional system displays bright single-photon emission as well as high stability at room temperature and is an excellent candidate for optomechanics.  The fast optical modulation of single-photon emission from BNNTs shows multiple electronic levels of the system and has potential applications in optical signal processing. 

\end{abstract}

\maketitle

Single-photon emitters (SPEs) are playing a key role in the route towards future quantum technologies such as quantum communication and quantum computing. Many solid-state SPEs have been explored, including quantum dots, carbon nanotubes, two-dimensional materials and color centers in crystals \cite{solidSPE}. They show potential in on-chip fabrication and integration which enable scalable structure for quantum information processing.   Recently, ultra-bright SPEs from point defects in 2D hexagonal boron nitride (hBN) with mono- and multilayers were discovered \cite{firsthBN}. As a well-suited platform for room-temperature quantum emission, 2D hBN has attracted a plethora of studies, including photoluminescence spectra distributed over a wide range \cite{multicolor,temperature,complex-level,photoinduced,attributes}, stable quantum emission at up to 800K \cite{robust800K,temperature}, and other properties \cite{nonlinear-excitation,firsthBN,defectstates-theory,firstprinciple,nonmagnetic}. 
Moreover, a recent work reported single photon emission from bundles of  boron nitride nanotubes (BNNTs) with an average diameter of 5 nm \cite{curved}. It reports bleaching under 532 nm excitation, and  notes the bleaching can be alleviated with  N,N'-Dimethylacetamide (DMAc) solutions and indium tin oxide (ITO) substrates \cite{curved}.

\begin{figure}[bh]
\includegraphics[width=\linewidth]{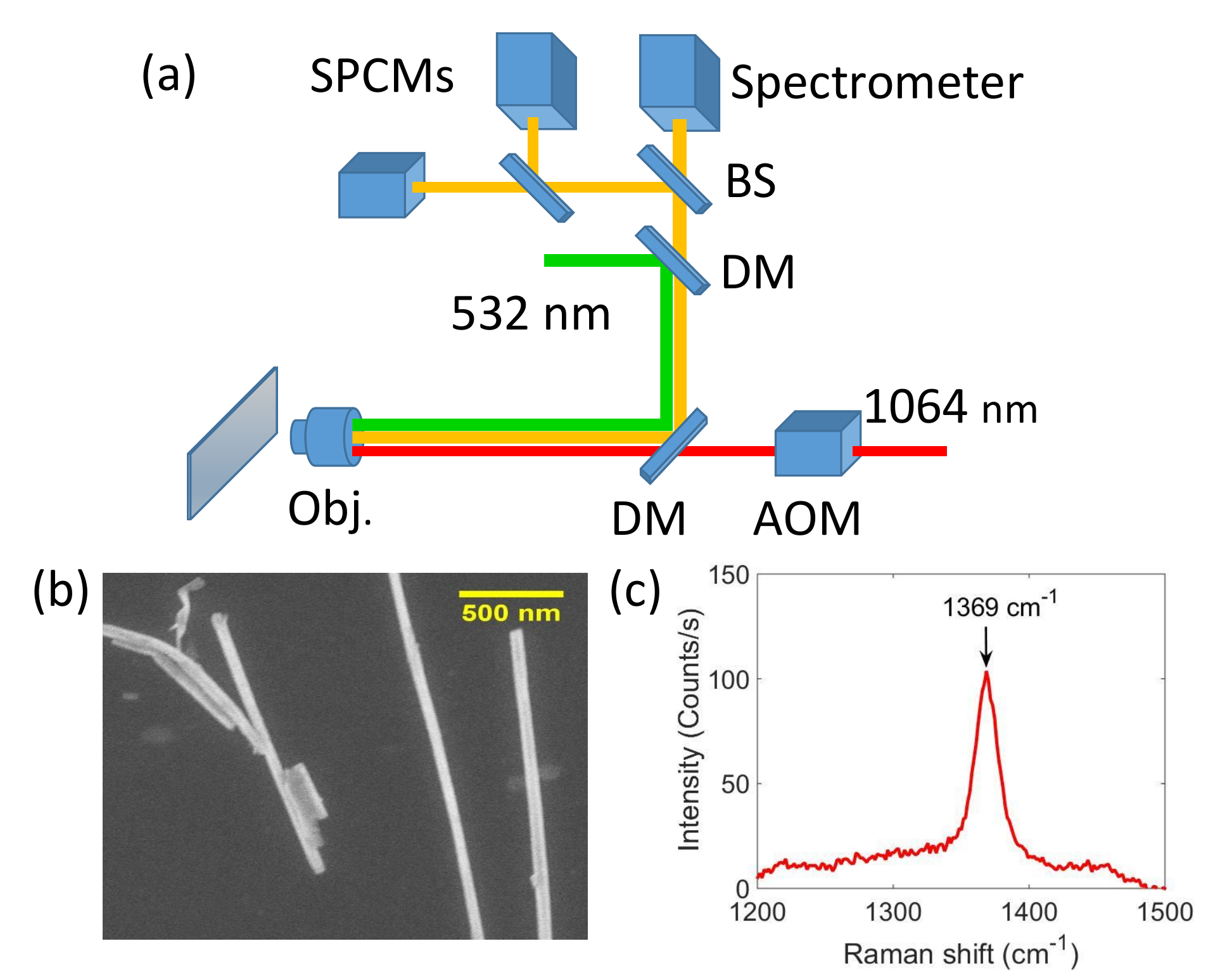}
\caption{(a) Simplified schematic of the experiment (SPCMs: single photon counting modules; BS: beam splitter; DM: dichroic mirror; Obj: objective lens; AOM: acousto-optic modulator). A green laser ($\lambda$ = 532 nm) is used to excite defects in BNNTs  on a glass slide. 
The emitted fluorescence (shown in orange) is collected with the objective lens (NA=0.9), and detected by two photon counters and a spectrometer. A near-infrared laser ($\lambda$ = 1064 nm) is used  for optical modulation. (b) A scanning electron microscope (SEM) image of the boron nitride nanotubes. The diameter of nanotubes is approximately 50 nm. (c) A Raman spectrum of a BNNT. The Raman shift is 1369 cm$^{-1}$.
}
\label{setup}
\end{figure}

\begin{figure*}[t]
\centerline{\includegraphics[width=\linewidth]{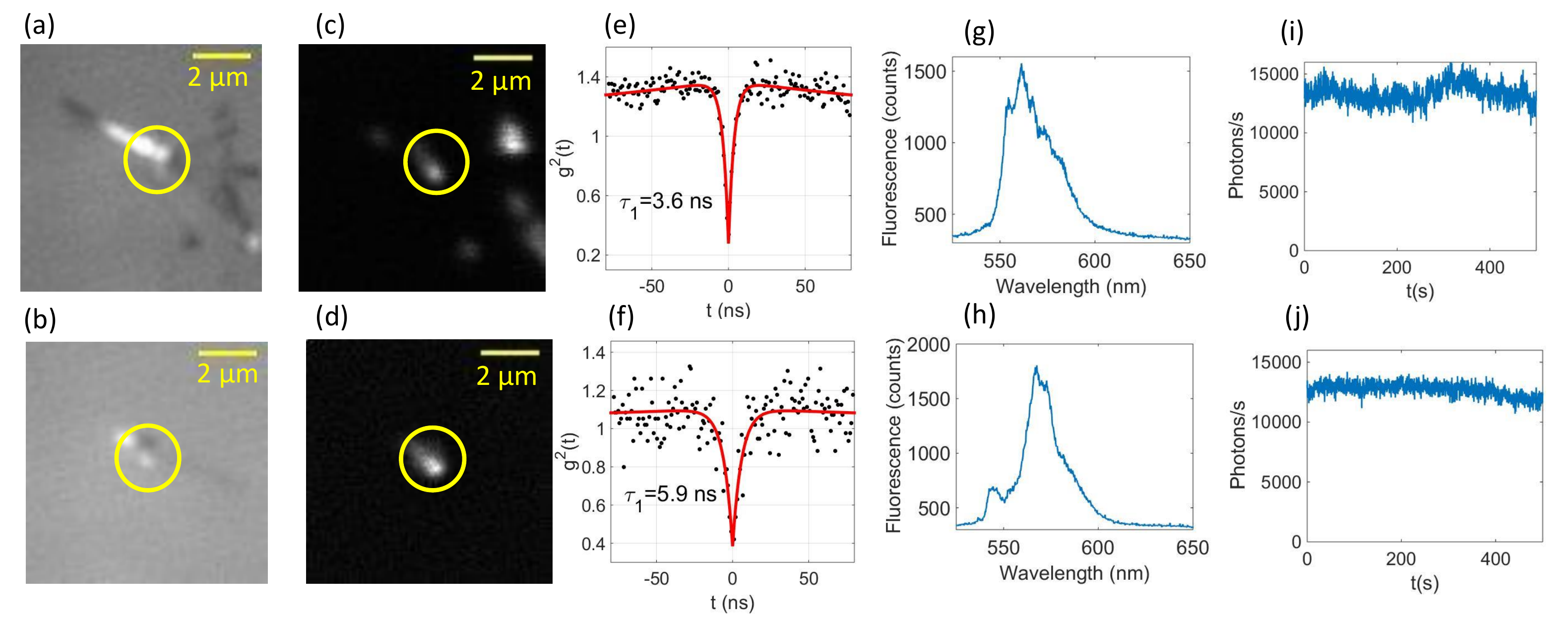}}
\caption{(a)-(b) Camera images under white light for two different defects in different BNNTs. Yellow circled regions indicate the positions of the defects. (c)-(d) Confocal photoluminescence map of the color centers under the 532 nm excitation. (e)-(f)) Measured second-order autocorrelation functions of the two color centers. $g^2(0)<0.5$ is satisfied in both cases, which proves that they are SPEs. (g)-(h) Measured spectral features of two color centers with a 550LP filter. (i)-(j) Photon emission rate as a function of time for 500s which demonstrates the stability of the point defects.}
\label{spectrum}
\end{figure*}

A BNNT can simply be seen as a rolled boron nitride sheet \cite{BNNT,BNNT2}. They are structurally similar to carbon nanotubes, but have wide band gaps (5.95 eV), and high thermal and chemical stabilities \cite{BNNT2}. With these features, BNNTs emerge as a natural candidate for nanotube-based nanodevices and protective shields, especially in hazardous and extreme environments \cite{BNNTapplication}.
Compared to 2D hBN, a BNNT will allow us to explore the effects of dimension and curvature on the  quantum emitters embedded inside. The optically active SPEs in low dimensions will also open up opportunities in quantum technologies with opto-mechanical resonators or other hybrid systems \cite{Resonator2,tunable,spinhBN,suspended,deterministic}.

In this work, we identify and study single photon emissions from point defects in BNNTs with an average diameter of 50 nm. Thanks to their intrinsic curvature, 50-nm-diameter BNNTs host abundant SPEs comparing to 2D hBN. We observe that their SPEs are much more stable than those in bundles of  5-nm-diameter BNNTs. 
Finally, we implement optical modulation of the fluorescence from the BNNTs by applying a near-infrared (NIR) laser with a wavelength of 1064 nm.

We use a home-built confocal microscope to selectively excite point defects in the 50 nm diameter BNNTs, as shown in Fig.\ref{setup}.(a). We dilute the BNNT powder in an ethanol solution and disperse it on a glass slide. The color centers are pumped to the excited state with a 300$\mu$W 532 nm continuous-wave (CW) laser focused through an air objective lens with an NA of 0.9. The fluorescence (shown in orange in Fig.\ref{setup}.(a)) is collected with the objective lens and is guided with beam splitters and dichroic mirrors to the single photon counters and a spectrometer. 
For the optical modulation of the photon emission, an additional 1064 nm laser (shown in red) is superimposed onto the 532 nm beam and the power is controlled with an acousto-optic modulator (AOM). Fig.\ref{setup}.(b) presents an SEM image of the BNNTs. The SEM image confirms that the nanotubes have an average diameter of approximately 50 nm and the nanotubes are well separated. Fig.\ref{setup}(c) presents the Raman scattering spectrum of the BNNTs. The Raman shift here is 1369 cm$^{-1}$ which agrees well with the results of earlier studies \cite{firsthBN,nonlinear-excitation,BNNT2,Raman,attributes}.

Fig.\ref{spectrum} displays different characteristics of the quantum emitters recorded from the BNNTs at room temperature. The camera images are collected from back reflection under white light and the results are displayed in Fig.\ref{spectrum}.(a)-(b). We notice that some parts of the nanotubes are arched, rather than being completely attached to the substrate surface. The nanotubes with a bent structure tend to have more defects. The confocal images (Fig.\ref{spectrum}.(c)-(d)) are taken by collecting the fluorescence from the defects excited with a 532 nm sub-bandgap pump laser.

\begin{figure*}[t]
\centerline{\includegraphics[width=0.8\linewidth]{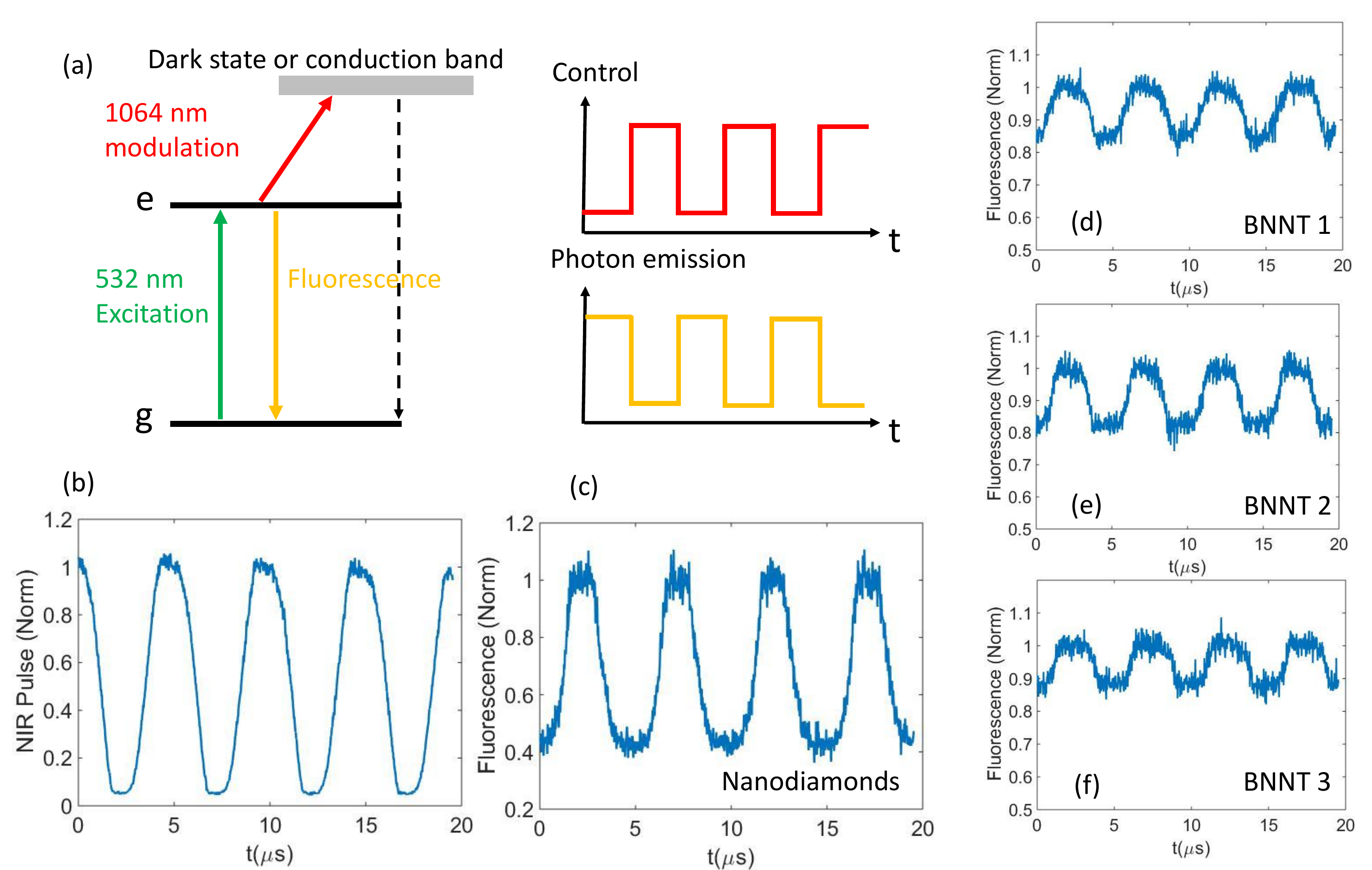}}
\caption{Optical modulation of photon emission from BNNTs under NIR illumination. (a) Schematic demonstration of the optical switching. The 1064 nm laser, acting as the control signal, is applied on top of the 532 nm beam. The orange line represents the fluorescence signal and it is modulated by the control signal. (b)The pulse shape of the 1064 nm laser. The signal is taken when only the 1064 nm beam is on. (c)By controlling the power of the 1064 nm laser with an AOM, we get the optical modulation of the fluorescence from NV centers in nanodiamonds. (d)-(f) The same condition is applied to the BNNTs, and a modulation of up to 20\% is achieved.}
\label{modulation}
\end{figure*}

To demonstrate that the collected signal in the confocal photoluminescence map comes from a SPE, we measure the second-order autocorrelation functions $g^2(\tau)$ using a Hanbury Brown and Twiss interferometry setup. The results are shown in Fig.\ref{spectrum}.(e)-(f). Both autocorrelation curves show $g^2(0)<0.5$, proving that both point defects are SPEs. The autocorrelation functions were fit by a three-level model \cite{firsthBN,g2function,threelevel},
\begin{equation}
g^2(\tau)=1-(1+a)e^{(-\left|{\tau}\right|/\tau_1)}+a e^{(-\left|{\tau}\right|/\tau_2)}
\end{equation}
where parameters $\tau_1$ and $\tau_2$ are the lifetimes of the excited state and the metastable state, respectively. $a$ is the fitting parameter. 
From the first emitter, we obtain $g^2(0)=0.28$ and the fluorescence lifetimes extracted from the fitting of the $g^2(\tau)$ curve are $\tau_1=3.6$ ns and $\tau_2=271.8$ ns.
The other  emitter has $g^2(0)=0.38$  while the fitting yield lifetimes of $\tau_1=5.9$ ns and $\tau_2=321.3$ ns.
The fluorescence lifetimes vary largely among different samples. This could be due to  local strains in the BNNTs. In addition, the diameter of BNNTs is much smaller than the wavelength of the excitation laser, which may cause nano-antenna effects that affects  the lifetimes \cite{curved,antennas}. Nevertheless, considering the uncertainty in the fittings, the measured lifetimes are  consistent with the results for SPEs in  hexagonal boron nitride monolayers \cite{firsthBN}.

For characterization of the defect, we record the photoluminescence spectra of each quantum emitter (presented in Fig.\ref{spectrum}.(g)-(h)). Both emitters have broad features and asymmetric line shapes in the spectra. These could be due to phonons in the nanotubes and background emissions. The peaks for the two emitters are found at 571 nm and 569 nm, respectively, presumably attributed to the zero phonon line (ZPL) of the defects. Based on the ZPL classification defined in \cite{multicolor}, the defects in our BNNTs belong to Group 1 and the fluorescence agrees well with the features of the spectra.

A previous study on bundles of BNNTs with  an average diameter of 5 nm reported obvious bleaching under 532 nm excitation and it used DMAc solutions and ITO substrates to resolve  bleaching \cite{curved}. 
On the contrary, in our experiment, we simply dilute the 50 nm BNNTs in an ethanol solution and disperse them on a glass slide. The single-photon emissions from our sample are  stable for over 20 minutes under the 532 nm excitation. Besides, we notice that some parts of the nanotubes are arched in air instead of being  adhered onto the substrate. Stable photon emissions are also observed in those arched sections. Therefore, in our sample, we surmise that the stability of the emission has little to do with the substrate material. 
To investigate the stability of the point-like defects in BNNTs, we maintain a constant pumping beam and track the photon number for 500 seconds as shown in Fig.\ref{spectrum}.(i)-(j). We find that  the photon numbers vary less than 20\% for 500s, which proves the stability of photon emissions\cite{firsthBN,robust800K,attributes}. 
For comparison, we have also conducted experiments on BNNT samples with an average diameter of about 5 nm under the same condition, on both ITO and glass substrates. It is much more difficult to find a SPE and the emissions are very unstable under the 532 nm illumination. Most of them bleach within 10 seconds. This is consistent with former results \cite{curved} and is likely due to a much larger curvature in 5-nm-diameter BNNTs. 

For further advancements in photonic devices and  quantum signal processing with SPEs, realization of  optical transistors has been pursued and implemented in several systems.  Previously, optical transistors have been realized in several systems including photonic crystals \cite{switchcavity}, optomechanical resonators \cite{switch-opto}, single molecules \cite{single-molecule} and single nitrogen-vacancy (NV) centers in nanodiamonds \cite{nvmodulation}. Similar to  NV centers in nanodiamonds, SPEs in hBN have been proved to be stable at room temperature, but their energy levels are still an open question. Several theories and calculations of possible energy levels are  available based on the measured spectra from quantum emitters \cite{firsthBN,multicolor,firstprinciple,complex-level,nonmagnetic}.
The optical modulation of SPEs in BNNTs with a NIR laser will not only create optical transistors but can also provide additional information about the possible energy levels of SPEs in BNNTs.

We perform the optical modulation of SPEs in the same 50 nm diameter BNNT sample deposited on a glass slide. An additional 1064 nm laser beam is superimposed onto the excitation laser for the modulation of the photon emission. Fig.\ref{modulation}.(a) shows the schematics for the optical modulation of BNNTs,  similar to the scheme for the NV centers of nanodiamonds in \cite{nvmodulation}. With a CW 532 nm pump laser, the point defect in BNNTs show stable fluorescence as shown in Fig.\ref{spectrum}.(i)-(j). Simultaneously, the non-resonant NIR laser is applied and drives the excited state to the dark state, which is brought back to the ground state through non-radiative decay. Therefore, the NIR beam serves as a control laser which reduces the fluorescence signal from the quantum emitters. Furthermore, controlled with an AOM, the NIR beam is modulated to a square pulse and the number of the emitted photons from the color center follows an opposite pattern. 

To begin with, we examine the pulse shape of the 1064 nm laser controlled with the AOM as shown in Fig.\ref{modulation}.(b). The total span of the recorded data is 20 $\mu$s while the modulated square pulse has a period of 5 $\mu$s. The pulse shape here is described by the number of photons collected from the back reflection and is normalized. Due to background noise, the measured minimum  amplitude for the  pulse is not zero.

Next, we investigate the effect of the NIR illumination on the fluorescence from the color centers in the BNNTs. For the excitation, a 532 nm pump beam with a constant power of 300 $\mu$W is used and the maximum power of the 1064 gating laser is 800 mW before the objective lens. The time-resolved fluorescence from the BNNTs is measured by continuous tracking of the photon numbers received at the single photon counter. With the modulated NIR laser and the constant green laser, we observe the obvious modulation of the fluorescence evolution in three different SPEs, which are demonstrated explicitly in Fig.\ref{modulation}.(d)-(f).

For comparison, we repeat the procedures above with NV centers in nanodiamonds with an average diameter of 40 nm. The power used for the green and NIR lasers are kept the same and is modulated with the same pulse shape as above. Fig.\ref{modulation}.(c) shows the optical modulation of NV centers in the nanodiamonds. In our system, we were able to achieve up to 60\% modulation of a NV center fluorescence, which is comparable to the 80\% modulation reported in \citep{nvmodulation}.  The BNNTs, on the other hand, showed an optical modulation of 10\% to 20\%. The lower modulation depth may be attributed to shorter lifetimes of SPEs in BNNTs \cite{multicolor,calculatedenergy}. The negative charge state (NV$^{-}$) in nanodiamonds has an average lifetime of about 20 ns \cite{NVlifetime} while the measured lifetime in BNNTs ranges from 3 ns to 5 ns. 
Therefore, the fluorescence process is much faster in BNNTs than that of NV centers in nanodiamonds.
Because of the shorter lifetime of the excited state $\ket{e}$ in BNNTs, it will be more difficult to pump it from $\ket{e}$ to the dark state with a NIR laser. 
Meanwhile, the quantum emission in BNNTs should have a shorter response time and accordingly a faster optical modulation. However, in our current experimental system, the  modulation speed is limited by the response time of our AOM, which is about 1$\mu$s. With improved experimental devices, the modulation limit of about 5 ns for the material should be achievable.

In conclusion, we investigate the quantum emission from boron nitride nanotubes. Results prove that BNNTs with a diameter of about 50 nm is a platform for stable SPEs at room temperature. The high quality of the quantum emission and the 1D nature of BNNTs provide potential applications in quantum technologies and hybrid optomechanics. More importantly, we achieve the fast optical modulation of the single photon emission from the BNNTs at room temperature. The ability to control the signal from a 1D single photon source will enable a foundation for single photon signal processing in integrated nanophotonic systems.

\section*{Funding}
NSF (PHY-1555035); Tellabs Foundation; Rolf Scharenberg Graduate Research Fellowship from Purdue University.


%

\end{document}